\documentclass[12pt]{article}
\usepackage{graphics,psfrag,amsbsy}
\usepackage[mathscr]{eucal}
\usepackage{euler}
\usepackage{amsmath}
\usepackage{amssymb}

\def\email#1{\date{\tt#1}}

\begin{document}

\title{Mathematical analysis of fully coupled
approach to creep damage}
\author{A.V. Shutov, A.-M. Saendig \\
\emph{Institute for Applied Analysis and Numerical Simulation} \\
\emph{70569 Stuttgart, Germany}}

\email{shutov@ngs.ru,saendig@ians.uni-stuttgart.de}

\maketitle \thispagestyle{empty}

\begin{abstract}
We prove the existence and uniqueness of solution
to a classical creep damage problem.
%Conventional creep damage model is analyzed.
We formulate a sufficient condition for the problem to have a
unique smooth solution, locally in time. This condition is stated
in terms of smoothness of given data, such as solid geometry,
boundary conditions, applied loads, and initial conditions.
Counterexamples with an arbitrary small lifetime of a structure
are also given, showing the mechanical interpretation of imposed
smoothness conditions. The proposed theory gives a rigorous
framework for a strain localization analysis.
The influence of the % damage as well as of the
damage gradient on the
strain localization process is characterized within this framework and
a measure of the damage localization is proposed.
\end{abstract}


\


\section{Introduction}

Structures made of metals and alloys are often used in different
brunches of industry at elevated temperatures (higher than 0.3
times the melting temperature). Typical examples are pressurized
pipes and vessels in power and chemical plants, gas turbines and
so on. Even subjected to moderate loads these structures
experience irreversible creep deformations which influence the
stress response in long time scales. The lifetime of such
structures is limited by damage processes induced by the
nucleation and growth of microscopic cracks and cavities. The finite
element method (FEM) is commonly used for numerical analysis of
nonlinear creep-damage response in the framework of continuum
damage mechanics (see \cite{Hayhurst1}) to estimate the remaining
lifetime. The classical creep models are described and analyzed in
the following monographs \cite{Rabotnov1}, \cite{Kachanov2},
\cite{Riedel1}, \cite{Skrzypek1}.

The creep behavior is divided into three stages. The initial stage
is characterized by hardening behavior with decreasing creep
strain rate. The second stage is the stationary creep with a
constant creep strain rate. The last stage is the tertiary creep
characterized by increasing creep strain rate and a dominant
softening of the material followed by a complete rupture. The most
popular constitutive law for the second stage was proposed by
Norton \cite{Norton} and postulates the stationary creep rate as
a power law function of the stress tensor. This constitutive law
is modified by use of time or hardening parameters (see, for example,
\cite{Skrzypek1}, \cite{Hayhurst4}) to take the primary creep into
account. A new internal continuity parameter $\psi$ was introduced
in the original work of Kachanov \cite{Kachanov1} to simulate the
material damage within the tertiary creep. This continuity
parameter is often replaced by a dual variable, namely, Rabotnov's
damage parameter $\omega=1-\psi$ (see \cite{Rabotnov1}). Within
Kachanov-Rabotnov's approach the damage rate is postulated as a
function of the stress, the temperature and the current damage
state. This is regarded as a foundation of continuum damage
mechanics (CDM).

One of unsolved problems of computational CDM is the spurious
mesh-dependence of FEM simulations (compare \cite{Liu1},
\cite{Murakami1}, \cite{Altenbach1}, \cite{Shutov1}, \cite{Shutov2})
which leads to physically unrealistic results.
Therefore numerous regularization techniques were proposed
to prevent this mesh-dependence
(see, for example, \cite{Saanouni1}, \cite{Hall1}, \cite{Murakami1}).

Proposed material models and regularization techniques are
generally tested by series of numerical experiments. At the same
time the mathematical treatment of nonlinear material models is
very poor. Some mathematical results are given in \cite{Deuring1},
\cite{Bonnetti1}, \cite{DeSimone1}, \cite{Mielke1}. In
\cite{Deuring1} the local existence and uniqueness is proved for a
coupled creep-damage model assuming the elastic properties are not
influenced by damage evolution (partly coupled approach).

For most of the used damage models it is not clear whether
the corresponding boundary value problems are well-posed.
We say that a given problem is well-posed
(see, for example, \cite{Evans}) if
the problem in fact has a solution;
this solution is unique; and the solution depends continuously on
the data given in the problem. In case of a creep-damage problem such
given data are solid geometry, boundary conditions, applied loads,
initial conditions, and material constants. A mathematically consistent
problem statement is necessary for justification of analytical
(see the paper \cite{Shutov3}) and numerical techniques. Particulary
it specifies how the difference between exact and approximate solution
can be measured and what kind of perturbations of given data are allowed.

The proper mathematical analysis of nonlinear damage
models is complicated by instabilities
due to loss of ellipticity of the corresponding differential
operator (compare \cite{Hill1}, \cite{Hayhurst2},
\cite{Deuring1} for example).
On the other hand,
bifurcation does not happen before the appearance
of completely damaged zone with $\omega=\omega^{\ast}$, where
$\omega^{\ast}$ is a critical damage value (see, for example, \cite{Liu1}).
The period of
time required by the structure to reach this state is called
crack growth initiation time $t^{\ast}$.
Therefore we prove existence and uniqueness
of solution in sufficiently small time interval
before failure initiation. On this time interval
the deformation is stable and
the problem can be posed correctly. Thus,
%we concern the
%crack initiation, but
%we analyzethe stage of damage evolution up to
the analysis of crack propagation lies beyond the scope of this article.
%, which is, in fact, more complicated problem.

The article is organized so that technical details
of the proof do not obscure the main points. First, we
introduce an initial boundary value problem for fully coupled
creep-damage model. In the following section we give basic definitions
of function spaces, that are necessary for the formulation of the main result.
In section 4 the main existence and uniqueness theorem is
formulated. One counterexample is provided, which illustrates
the effect of damage localization. Finally we prove the main result and
summarize our main conclusions.

\section{Constitutive equations}

Let $\Omega \subset \mathbb{R}^2$ be a bounded domain which represents the solid.
In this work we confine ourselves to the plane-stress two-dimensional case.
But the theory proposed here can be easily generalized to three dimensions.
Let us assume a stationary temperature field. Therefore the constitutive
equations do not depend on temperature.

\subsection{Fully coupled damage model}

Suppose that the damage evolution is controlled by
the von Mises equivalent stress.
Accordingly to the classical Kachanov-Rabotnov concept
the constitutive equations for secondary and tertiary
creep are summarized as follows

\begin{equation}\label{consteq1}
\boldsymbol\sigma= \boldsymbol{C}^{\omega}(\boldsymbol\varepsilon -
{\boldsymbol\varepsilon}^{cr}) \quad \text{in}  \ \Omega \times [0,T],
\end{equation}
\begin{equation}\label{consteq2}
\dot{\boldsymbol\varepsilon}^{cr}= \frac{3}{2} A  \ \boldsymbol s \
(\sigma_{vM})^{n-1} (1-\omega)^{-n} \quad \text{in}
 \ \Omega \times (0,T),
\end{equation}
\begin{equation}\label{consteq3}
\dot{\omega}= B  \ (\sigma_{vM})^{m} (1-\omega)^{-q}
\quad \text{in}  \ \Omega \times (0,T),
\end{equation}
\begin{equation}
\boldsymbol s = \boldsymbol \sigma - \frac{1}{3}tr(\boldsymbol \sigma)
\boldsymbol I, \quad
\sigma_{vM}=\sqrt{\frac{3}{2}\boldsymbol s : \boldsymbol s},
\end{equation}
where $\boldsymbol\sigma$ is the stress tensor,
$\boldsymbol{C}^{\omega}$ is the fourth-rank tensor
depending
on Rabotnov's damage parameter $\omega$,
 $\dot{( \ )}$ is the time derivative, $\boldsymbol{\varepsilon}^{cr}$ is
the creep strain,
$\boldsymbol s$ is the stress
deviator, $\sigma_{vM}$ is the von Mises equivalent stress, $\boldsymbol
I$ is the second rank unit tensor, and $A$, $B$, $n$, $m$, $q$ are material
constants. The influence of the damage
on elastic properties is given by the equation (see \cite{Lemaitre}, \cite{Liu1})
\begin{equation}\label{fulcup}
\boldsymbol{C}^{\omega}= \boldsymbol{C} (1-\omega).
\end{equation}
Here
$\boldsymbol{C}$ denotes the tensor of linear
elasticity of undamaged solid. $\boldsymbol{C}$ is linear, symmetric,
positive definite mapping.

Furthermore, we consider equilibrium equations
\begin{equation}\label{equilInvar}
 \boldsymbol \nabla \cdot \boldsymbol \sigma = - \boldsymbol q
 \quad \text{in}  \ \Omega \times [0,T],
\end{equation}
and strain-displacement relations
\begin{equation}\label{str-dispInvar}
\boldsymbol \varepsilon =\frac{1}{2}  \left(  \boldsymbol
\nabla \boldsymbol u + \boldsymbol \nabla
\boldsymbol u^T \right),
\end{equation}
where $\boldsymbol \varepsilon$ is the linearized strain tensor
and $\boldsymbol u$ is the displacement vector.
The quantities
$(\boldsymbol u, \boldsymbol\varepsilon^{cr}, \omega, \boldsymbol \sigma)$
depend on the space variable $x \in \Omega$ and the time
variable $t \in [0,T]$ for some $T > 0$.
The system is completed by boundary and initial conditions
\begin{equation}\label{boundc}
\boldsymbol u= \boldsymbol u^{\ast} \quad \text{on} \ \partial \Omega  \times [0,T],
\end{equation}
\begin{equation}\label{inicond}
{\boldsymbol\varepsilon}^{cr}|_{t=0} = {\boldsymbol\varepsilon}_0^{cr}
\quad \text{in} \  \Omega,
\end{equation}
\begin{equation}\label{inicond2}
{\omega}|_{t=0} = {\omega}_0 \quad \text{in} \  \Omega.
\end{equation}

\emph{Remark.}
The constitutive equation for damage evolution \eqref{consteq3}
was generalized by Hayhurst (see \cite{Hayhurst2}) so that $\dot{\omega}$
depends on the combination $\alpha \sigma_1 + (1-\alpha)\sigma_{vM}$ of the
maximal principal stress and von Mises equivalent stress.
But the proof of the main result (\emph{Theorem 4.1}) is essentially
based on the smoothness of constitutive equations, therefore we do not
analyze this popular model here.

\subsection{Remark on the partly coupled damage model}

Equation \eqref{fulcup} of the fully coupled model
is based on Kachanov's concept
of reduction of the effective load carrying area.
This equation is a fundamental form of elastic-damage coupling.

Within the partly coupled approach the influence of damage on
the elastic properties is neglected and
equation \eqref{fulcup} is replaced by
\begin{equation}\label{parcup}
\boldsymbol{C}^{\omega}=  \begin{cases}
    \boldsymbol{C} \quad \text{if} \ \omega < \omega^{\ast} \\
    \boldsymbol{0} \quad \text{if} \ \omega = \omega^{\ast}
  \end{cases},
\end{equation}
where $\omega^{\ast}$ stands for critical damage state.

In most of engineering applications this approach
is used as a simplified variant of the fully coupled relations in
order to decrease the computational effort.
Unlike the fully coupled model, the partly coupled approach does not
require a modification and decomposition of the stiffness matrix
on each time or iteration step.
As it was observed in \cite{Liu1} and \cite{Altenbach1},
the partly coupled approach
gives a good estimation of failure time for
some specimens and initial conditions.
Nevertheless, as it will be shown later, this
simplification should be used carefully.
% even for estimation of failure time.
This model does not take into account the
stress concentrations caused by damage inhomogeneity.
For instance, if $A=0$ in \eqref{consteq2}
and $B \neq 0$ in \eqref{consteq3}, then
the partly coupled system \eqref{parcup}
describes linear elasticity of homogeneous solid.

%The fully coupled approach predicts zero stiffness in the
%critical damage state $\omega=1$,

Existence and uniqueness for partly coupled model were proved
 in \cite{Deuring1} in case of thin-walled
structures. Thus, the plane stress was covered as a
special case of shell geometry.
We generalize the existence proof, given in
\cite{Deuring1}, to take the fully coupled
damage model into account.

\section{Basic notations}

The creep-damage problem \eqref{consteq1} --- \eqref{inicond2}
can be formulated in a well-posed manner with the help of
suitable function spaces.
Field variables which describe the structure
are considered to be elements of these infinite dimensional spaces.
The corresponding function norms should
take into account the physical essence of the
problem and the properties of the system of equations.
%In that case the norms are used to give a clear definition of terms
%"stable", "unstable", "convergency", and so on.

\subsection{Definition of function spaces}

%For $n \in \mathbb{N}$, $x \in \mathbb{R}^n$ we define $l_1$-norm of $x$ by $|x|_1=|x_1|+...+|x_n|$.
Let $B$ be a Banach space endowed with a norm $\|\cdot\|_B$
and $T$ be a positive real number.
We introduce a space of
continuous $B$-valued functions defined on the interval $[0,T]$.

\textbf{Definition 1}
\begin{equation}\label{def1}
C^0([0,T],B):=\{\varphi: [0,T] \rightarrow B, \ \varphi \
\text{is continuous}\}.
\end{equation}
 This space is a Banach space equipped with the norm
\begin{equation}\label{def11}
\|u\|_{B,\infty}=sup\{\|u(t)\|_B: t \in [0,T]\}.
\end{equation}

Let $k \in \mathbb{N}_0$ and $p \geqslant 1$. We define the usual
Sobolev space $W^{k,p}(\Omega)$
 (see, for example, \cite{Adams}, \cite{Evans}, \cite{Gilbarg}).

\textbf{Definition 2}
\begin{equation}\label{def2}
W^{k,p}(\Omega):=\{u \in L_p: D^{\alpha} u \in L_p \ \text{for all}
\ \alpha=(\alpha_1, \alpha_2)\in \mathbb{N}^2_0,
\alpha_1 + \alpha_2 \leq k \}
\end{equation}
endowed with the norm
\begin{equation}\label{def12}
\|u\|_{k,p}=\big( \sum\limits_{\alpha \in \mathbb{N}^2_0, \ |\alpha| \leq k}
\|D^{\alpha} u\|^p_p \big)^{\frac{1}{p}}.
\end{equation}
 Here $D^{\alpha} u$ are generalized derivatives
of the order $| \alpha |= \alpha_1+\alpha_2$.

Beside the Sobolev space $W^{k,p}$ we will need a proper subspace
$W^{k,p}_0(\Omega) \subset W^{k,p}$ which is defined as follows.

\textbf{Definition 3}

Let $C^{\infty}_0(\Omega):=\{ \varphi \in  C^{\infty}(\Omega),
\text{supp} \ \varphi \subset \Omega\}$
be the set of smooth functions that vanish near the boundary
$\partial \Omega$. Then
\begin{equation}\label{def3}
W^{k,p}_0(\Omega):= \overline{C^{\infty}_0(\Omega)}^{W^{k,p}(\Omega)}
\end{equation}
is the closure of $C^{\infty}_0(\Omega)$ with respect to the norm $\|u\|_{k,p}$.
Note that functions from $W^{k,p}_0(\Omega)$ vanish on the boundary in
the trace sense (see \emph{Definition 4}).

\textbf{Theorem 3.1} (Imbedding theorem, see Theorem 7.26 in \cite{Gilbarg})

\emph{Let $p>2$ and $\Omega$ be a Lipschitz domain
in $\mathbb{R}^2$. Then
$W^{k,p}(\Omega)$ is continuously imbedded in $C^{0,k-2/p}(\overline{\Omega})$.}

\textbf{Corollary 3.1} (Sobolev inequality)

\emph{Let $p>2$ and $\Omega$ be a Lipschitz domain in $\mathbb{R}^2$.
Then there is a
constant $C_I< \infty$ with}
\begin{equation}\label{Imbed}
\|u\|_{C^{0}(\overline{\Omega})} \leq C_I \|u\|_{W^{1,p}(\Omega)}.
\end{equation}

Furthermore we need the traces of functions from $W^{2,p}(\Omega)$
on the boundary $\partial \Omega$.
% and vise versa the extension
%of the functions defined on $\partial \Omega$ into the domain $\Omega$.

\textbf{Definition 4}

Let $\Omega$ be a bounded domain with $C^{1,1}$-boundary.
That means the boundary $\partial \Omega$ is locally given by a function
with a Lipschitz continuous derivative. Suppose $p > 1$. Then the trace space
of functions from $W^{2,p}(\Omega)$ is defined as (see \cite{Grisvard}, pp. 37-38)
\begin{multline}\label{TrSp}
W^{2-\frac{1}{p},p}(\partial \Omega)=
\{u \in W^{1,p}(\partial \Omega):
\int\limits_{\partial \Omega} \int\limits_{\partial \Omega}
\frac{| D^{\alpha} u (x) - D^{\alpha} u (y)|^p}
{|x-y|^p} ds_x ds_y < \infty
 \\ \text{for all} \ \alpha=(\alpha_1, \alpha_2)\in \mathbb{N}^2_0,
\alpha_1 + \alpha_2 \leq 1 \}.
\end{multline}
If $p > 1$, then the trace operator
\begin{equation}\label{TrOp1}
Tr: W^{2,p}(\Omega) \rightarrow W^{2-\frac{1}{p},p}(\partial \Omega),
\end{equation}
\begin{equation}\label{TrOp2}
Tr: u \mapsto u|_{\partial \Omega}
\end{equation}
is well defined in the classical sense.

%Then for $s \in (0,2)$
%we designate the Sobolev trace space of fractional order $s$ and exponent $p$
%trough $W^{s,p}(\partial \Omega)$.

%Field variables which describe the structure
%should belong to the proper classes.
If the time $t$ is fixed, then
we consider the field variables to be the functions, which are defined in $\Omega$
and belong to the proper function spaces.
We will use the following abbreviations of function spaces and subsets:
\begin{itemize}
\item $X_p:=(L_p(\Omega))^2$ for the volumetric loads,
\item $Y_p:=W^{1,p}(\Omega)$ for the components of creep strain tensor,
\item $Y^4_p:=(Y_p)^4$ for the creep strain tensors,
\item $V_p:=(W^{2,p}(\Omega))^2$ for the displacement fields,
\item $V^0_p:=(W^{2,p}(\Omega) \cap W^{1,p}_0 (\Omega))^2$
for the displacement fields with
a vanishing boundary values (which correspond
to the solid clamped at the boundary),
\item $Y^{\beta_1, \beta_2}_p:=
\{\omega \in  Y_p:
0 \leq \omega(x) \leq 1- \beta_1, \|\omega\|_{Y_p} \leq \beta_2\}$
for the damage fields,
where $\frac{1}{2}>\beta_1>0, \beta_2>0$ are fixed constants.
Accordingly to \emph{Corollary 3.1}, $\omega(x)$ is well
defined and $Y^{\beta_1, \beta_2}_p$
is a closed subset of $Y_p$.
\end{itemize}

\emph{Remark.}
The condition $0 \leq \omega(x) \leq 1- \beta_1$ is natural
to guarantee that the elasticity tensor \eqref{fulcup} is
positive definite. The second condition
$\|\omega\|_{Y_p} \leq \beta_2$ imposes additional constraints
both on the damage field and on the damage gradient.
%  is used to get a uniform estimate
%of moduli of continuity of $\omega$.

\subsection{Reduction to zero prescribed displacements}
In this subsection we reduce the boundary value problem
\eqref{consteq1} --- \eqref{inicond2}
to the case of zero prescribed displacements along the
boundary $\partial \Omega$.

\textbf{Theorem 3.2}

\emph{ Suppose that displacements, which are given on the boundary,
satisfy the following smoothness condition:
$\boldsymbol{u}^{\ast} \in (W^{2-\frac{1}{p},p}(\partial \Omega))^2$.
Then there is a function $\boldsymbol{\hat{u}} \in V_p$
with $\hat{\boldsymbol{u}}|_{\partial \Omega}=
\boldsymbol{u}^{\ast}$ in the trace sense} (see \cite{Grisvard}).

In what follows, we designate $\boldsymbol{\hat{u}}$
by the same symbol as $\boldsymbol{u}^{\ast}$.

Now we reformulate our problem in a standard way. We search a vector
$(\boldsymbol u, \boldsymbol{\varepsilon}^{cr}, \omega)
\in V^0_p \times Y^4_p \times Y_p$, such that
$(\boldsymbol u + \boldsymbol u^{\ast},
\boldsymbol{\varepsilon}^{cr}, \omega) \in V_p \times Y^4_p \times Y_p$
is a solution of \eqref{consteq1} --- \eqref{inicond2}.

\subsection{Compact form of evolution equations}

In this subsection we rewrite the evolution equations
\eqref{consteq2}, \eqref{consteq3}
in a compact form
%under plane stress conditions.
\begin{equation}\label{comact1}
\dot{\boldsymbol\varepsilon}^{cr}(x,t)=\mathcal{R}(\rho
( \boldsymbol u, {\boldsymbol\varepsilon}^{cr}, \omega)(x,t))
\end{equation}
\begin{equation}\label{comact2}
\dot{\omega}(x,t)=\mathcal{S}(\rho
( \boldsymbol u, {\boldsymbol\varepsilon}^{cr}, \omega)(x,t)).
\end{equation}
To this end we introduce for every
$(\boldsymbol u, {\boldsymbol\varepsilon}^{cr}, \omega)
\in V_p \times Y^4_p \times Y^{\beta_1, \beta_2}_p$,
%${\boldsymbol\varepsilon}^{cr} \in Y^4_p $, and
%$\omega \in Y^{\beta_1, \beta_2}_p$
%\begin{equation}\label{comact3}
%\rho( \boldsymbol u, {\boldsymbol\varepsilon}^{cr}, \omega) \in Y^7_p, \
%\end{equation}
\begin{equation}\label{comact22}
\rho( \boldsymbol u, {\boldsymbol\varepsilon}^{cr}, \omega) :=
\big( \varepsilon_{11}(\boldsymbol u), \varepsilon_{22}
(\boldsymbol u), \varepsilon_{12}(\boldsymbol u),
 \varepsilon^{cr}_{11}, \varepsilon^{cr}_{22},
 \varepsilon^{cr}_{12}, \omega \big) \in Y^7_p,
\end{equation}
\begin{equation}\label{comact3}
\boldsymbol\varepsilon (\boldsymbol u):= \frac{1}{2}  \left(  \boldsymbol
\nabla \boldsymbol u + \boldsymbol \nabla
\boldsymbol u^T \right) \in Y^4_p.
\end{equation}
For every $\rho \in \mathbb{R}^7$ we define
\begin{equation}\label{comact4}
\mathcal{R}(\rho):= \frac{3}{2} A  \ \boldsymbol s (\rho) \
(\sigma_{vM}(\rho))^{n-1} (1-\rho_7)^{-n},
\end{equation}
\begin{equation}\label{comact5}
\mathcal{S}(\rho):= B  \ (\sigma_{vM}(\rho))^{m} (1-\rho_7)^{-q},
\end{equation}
\begin{equation}\label{comact6}
\sigma_{vM}(\rho):= P (\sigma_{11}(\rho), \sigma_{22}(\rho),
\sigma_{12}(\rho)),
\end{equation}
\begin{equation}\label{comact7}
\boldsymbol s (\rho):= \boldsymbol \sigma (\rho) -
\frac{1}{3}tr(\boldsymbol \sigma (\rho)) \boldsymbol I,
\end{equation}
\begin{equation}\label{comact8}
\boldsymbol \sigma (\rho) := \left(\begin{array}{cc}
       \sigma_{11} & \sigma_{12}  \\
       \sigma_{12} & \sigma_{22}  \\
\end{array} \right)(\rho),
\end{equation}

\begin{equation}\label{comact9}
\left(\begin{array}{c}
       \sigma_{11}   \\
       \sigma_{22}   \\
       \sigma_{12}   \\
\end{array} \right)(\rho) := (1-\rho_7) \frac{E}{1-\nu^2}
\left( \begin{array}{ccc}
       1   & \nu  &  0 \\
       \nu &  1   &  0 \\
       0   &  0   &    \frac{1-\nu}{2}  \\
\end{array}  \right)
\left(\begin{array}{c}
       \rho_1-\rho_4   \\
       \rho_2-\rho_5   \\
       2(\rho_3-\rho_6)   \\
\end{array} \right),
\end{equation}
\begin{equation}\label{comact10}
P(z_1,z_2,z_3):= \sqrt{z^2_1+z^2_2-z_1 z_1 + 3 z^2_3}
\quad \text{for every} \quad z \in \mathbb{R}^3.
\end{equation}

The constitutive relations \eqref{comact9},
\eqref{comact10} are obtained from the
general 3D relations under plane stress assumption
($\sigma_{13}=\sigma_{23}=\sigma_{33}=0$).

\section{Main result}

The existence and uniqueness theorem
states that a unique smooth solution
to the initial boundary value problem
\eqref{consteq1} --- \eqref{inicond2}
exists in a certain time interval.

\subsection{Formulation of the main theorem}

\textbf{Theorem 4.1}

\emph{ Let $\Omega$ be a bounded domain with $C^{1,1}$-boundary, $p>2$,
$T > 0$, $\boldsymbol q \in C^0([0,T],X_p)$, $\boldsymbol u^{\ast} \in V_p$,
${\boldsymbol\varepsilon}_0^{cr} \in Y^4_p$, and
$\omega_0 \in Y^{\beta_1, \beta_2}_p$.
Then there exists $T_1 \in (0, T]$ such that for any
$T' \in (0, T_1]$ there is a uniquely determined
mapping $(\boldsymbol u, \boldsymbol{\varepsilon}^{cr}, \omega) \in
C^0([0,T'],V^0_p \times Y^4_p \times Y_p)$ such that}
\begin{equation}\label{result2}
\boldsymbol \nabla \cdot \big((1-\omega) \boldsymbol{C}
(\boldsymbol\varepsilon (\boldsymbol u + \boldsymbol u^{\ast})
- {\boldsymbol\varepsilon}^{cr})\big) = - \boldsymbol q(t)
\quad \text{for all} \quad t \in [0,T'],
\end{equation}
\begin{equation}\label{result3}
(\boldsymbol{\varepsilon}^{cr}, \omega)(t)=
(\boldsymbol{\varepsilon}^{cr}_0, \omega_0)+
\int\limits_0^t \big(\mathcal{R}(\rho
( \boldsymbol u + \boldsymbol u^{\ast},
{\boldsymbol\varepsilon}^{cr}, \omega),
\mathcal{S}(\rho
( \boldsymbol u + \boldsymbol u^{\ast},
{\boldsymbol\varepsilon}^{cr}, \omega)\big)(s)ds
\end{equation}
\emph{for every $t \in [0,T']$. Here the evolution operators
$\mathcal{R}, \mathcal{S}$ are defined by
\eqref{comact1} --- \eqref{comact10}.}

\emph{Moreover,}
\begin{equation}\label{result1}
\omega(x,t)<1 \quad \text{for all} \quad (x,t) \in \Omega \times [0,T'],
\end{equation}
\begin{equation}\label{result4}
(\boldsymbol{\varepsilon}^{cr}, \omega) \in
C^1([0,T'], Y^4_p \times Y_p),
\end{equation}
\begin{equation}\label{result5}
\dot{\boldsymbol{\varepsilon}}^{cr}=\mathcal{R}(\rho
( \boldsymbol u + \boldsymbol u^{\ast},
{\boldsymbol\varepsilon}^{cr}, \omega), \quad
\dot{\omega}=\mathcal{S}(\rho
( \boldsymbol u + \boldsymbol u^{\ast},
{\boldsymbol\varepsilon}^{cr}, \omega), \quad
\end{equation}
\begin{equation}\label{result6}
{\boldsymbol\varepsilon}^{cr}(0) = {\boldsymbol\varepsilon}_0^{cr}, \quad
\omega(0) = \omega_0,
\end{equation}
\begin{equation}\label{result7}
\|{\boldsymbol\varepsilon}^{cr}(t)-{\boldsymbol\varepsilon}_0^{cr}\|_{Y^4_p}+
\|\omega(t)-\omega_0\|_{Y_p} \leq \min \big(\frac{\beta_1}{2(1+C_I)}, \frac{\beta_2}{2} \big)
\ \forall \ t \in [0, T'].
\end{equation}

\emph{Theorem 4.1} is proved in the next section.

\textbf{Corollary 4.1}

\emph{If the solid geometry, applied loads, prescribed displacements, and initial data
are smooth, then the fully coupled creep-damage model predicts a nonzero
lifetime $t^{\ast}$ of the structure with a lower estimate $T_1$ from Theorem 4.1
($t^{\ast}\geq T_1$).}

\emph{Remark.} \emph{Theorem 4.1} assures that $t^{\ast} \geq T_1 > 0$ only
for smooth domains without notches.
Numerous examples of FEM simulation
of notched specimens show that the predicted crack initiation
time $t^{\ast}$ tends to zero as the mesh-size decreases (\cite{Liu1}).

\emph{Remark.}
The lifetime estimate $T_1$ depends %only on the solid
%geometry, applied loads, prescribed displacements $\boldsymbol u^{\ast}$, initial
%creep strains ${\boldsymbol\varepsilon}_0^{cr}$, and
on the constants $\beta_1, \beta_2$. Moreover, as it will be clear from
the proof of \emph{Theorem 4.1}, $T_1=T_1(\beta_1/ \beta_2)$.
It is natural that $T_1 \rightarrow 0$ as $\beta_1 \rightarrow 0$ since
the lifetime of the structure made of almost broken
material ($\min\limits_{x \in \overline{\Omega}}(1-\omega_0)\rightarrow 0$)
is negligibly small.
Furthermore, $T_1$ tends also to zero as $\beta_2$ tends
to infinity even if $\beta_1$ is finite.
The physical interpretation of this result could be the following.
The rupture time can be negligibly small in the case of big gradients
of damage ($\| \omega_0 \|_{Y_p} \rightarrow \infty$)
even if the initial damage itself was not substantial
($\min\limits_{x \in \overline{\Omega}}(1-\omega_0) \sim 1$).

Example is provided in the subsections 4.2 showing that the dependence
of $t^{\ast}$ on $\beta_2$ can be interpreted as
 lifetime reduction due to damage localization.

\subsection{Counterexample: lifetime reduction due to local imperfections}

\begin{figure}\centering
\psfrag{L}[m][][1][0]{L} \psfrag{h}[m][][1][0]{h}
\scalebox{1}{\includegraphics{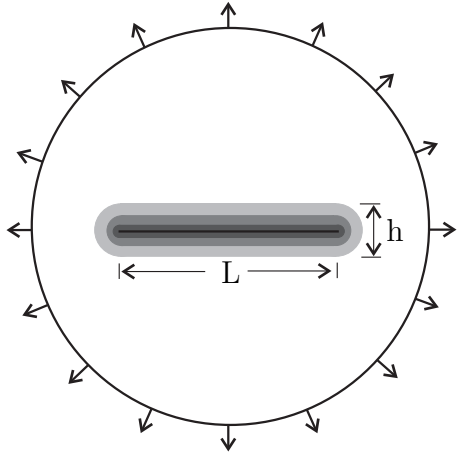}} \caption{System
configuration, boundary conditions, and initial damage \label{fig1}}
\end{figure}

Consider a solid loaded by prescribed displacements on it's boundary
as shown on figure \ref{fig1}. Assume that the boundary and prescribed
displacements are smooth.
We set the creep strain rate
to zero ($B=0$ in \eqref{consteq3}).
Consider a curve $l$ of length $L$ within the solid. Suppose that
initial damage is concentrated near the curve $l$
(see fig. \ref{fig1}) and the initial creep is zero
\begin{equation}\label{imperfect}
\omega_0(x):=\max(0, \frac{h - \text{dist}(x,l)}{2 h}),
\quad
{\boldsymbol\varepsilon}_0^{cr}:= \boldsymbol 0.
\end{equation}
It is obvious that
\begin{equation}\label{imperfect2}
\min(1-\omega_0) \equiv 1/2,   \quad  \| \omega_0 \|_{Y_p}
\rightarrow \infty \quad \text{as} \quad h \rightarrow 0.
\end{equation}

We assert that
\begin{equation}\label{imperfect3}
t^{\ast} \rightarrow 0 \quad \text{as} \quad h \rightarrow 0.
\end{equation}
To prove this assertion we can use the same argumentation
as used in \cite{Liu1}. The main reason the
lifetime is decreasing is because
%the initial stress state is given by
%elastic solution. Consequently
the stress concentration factor near
the curve tip tends to infinity as $h \rightarrow 0$.

%\subsection{Discussion of theoretical results}

%The counterexample 4.2.1 is based on the idea of using the damage localization process.

%In this subsection we discuss the requirements of \emph{Theorem 4.1}.
%We show that they have a physical meaning and the violation of one of them
%^directly affects the lifetime estimate.

\section{Proof of \emph{Theorem 4.1}}

To prove \emph{Theorem 4.1} we need several lemmas.

\subsection{Equilibrium equations with respect to $\boldsymbol u$ with a given
$\boldsymbol\varepsilon^{cr}$ and $\omega$}

Define the family $\big\{ \mathcal{L}_{\omega}\big\}_{\omega \in Y^{\frac{\beta_1}{2}, 2 \beta_2}_p}$
of operators of linear elasticity

\begin{equation}\label{linealst}
\mathcal{L}_{\omega}: V_p \rightarrow X_p
\end{equation}
by the rule
\begin{equation}\label{linealst2}
\mathcal{L}_{\omega}(\boldsymbol u):=\boldsymbol \nabla \cdot \big((1-\omega) \boldsymbol{C}
\boldsymbol\varepsilon (\boldsymbol u) \big)
\end{equation}
for $\omega \in Y^{\frac{\beta_1}{2}, 2 \beta_2}_p, \ \boldsymbol u \in V_p$.

\textbf{Lemma 5.1}

The operator
$\mathcal{L}_{\omega}$ is bounded for all $\omega
\in Y^{\frac{\beta_1}{2}, 2 \beta_2}_p$. Moreover,
the problem
\begin{equation}\label{probb}
\mathcal{L}_{\omega}(\boldsymbol u) = - \boldsymbol q
\end{equation}
 has
a unique solution $\boldsymbol u \in V^0_p$ for all
$\omega \in Y^{\frac{\beta_1}{2}, 2 \beta_2}_p, \ \boldsymbol q \in X_p$ and
\begin{equation}\label{linealst3}
\| \boldsymbol u \|_{V_p} \leq C_{5.1} \| \boldsymbol q \|_{X_p}.
\end{equation}

Here $C_{5.1} < \infty$ does not depend on $\boldsymbol q$.

\textbf{Proof.} The boundness of $\mathcal{L}_{\omega}$ follows from the following
computations
\begin{multline}\label{lemma5.1}
\| \mathcal{L}_{\omega}(\boldsymbol u) \|_{X_p}=
\| \boldsymbol \nabla \cdot \big((1-\omega) \boldsymbol{C}
\boldsymbol\varepsilon (\boldsymbol u) \big) \|_{X_p} \\
\leq
\| \boldsymbol \nabla \omega \|_{X_p} \cdot
\| \boldsymbol{C}
\boldsymbol \varepsilon (\boldsymbol u) \|_{C^0} +
\| 1- \omega \|_{C^0} \cdot
\| \boldsymbol \nabla \cdot \big( \boldsymbol{C}
\boldsymbol\varepsilon (\boldsymbol u) \big) \|_{X_p} \\
\leq
C(\beta_1, \beta_2) \| \boldsymbol u \|_{V_p}.
\end{multline}
Up to the rest of the article the expression $Q_1 \leq C \cdot Q_2$ should be
understood as follows. The quantities $Q_1$ and $Q_2$ are related to each other in
such a way that there is a suitable constant $C < \infty$, which depends only on
$( \Omega, E, \nu, p, n, m, q, \beta_1, \beta_2)$ and $Q_1 \leq C \cdot Q_2$.

For the proof of solvability of \eqref{probb} and for estimate \eqref{linealst3}
see \cite{Gilbarg} (p. 241). Particulary, we have the following inequality
\begin{equation}\label{damloc}
\| \boldsymbol u \|_{V_p} \leq C( \Omega, E, \nu, p)
\frac{\beta_2}{\beta_1} \| \boldsymbol q \|_{X_p}.
\end{equation}
The lemma is proved $\blacksquare$

\emph{Remark.}
The influence of the damage localization on the strain localization
is taken into account by \eqref{damloc}. Indeed,
\begin{equation}\label{localization}
C_{5.1} \rightarrow \infty, \quad \text{as} \ \frac{\beta_2}{\beta_1} \rightarrow \infty.
\end{equation}

We denote by $\mathcal{L}_\omega|_{V^0_p}$ the
restriction of $\mathcal{L}_\omega$ to $V^0_p$.
Let $\mathcal{L}^{-1}_\omega$ be the inverse to
$\mathcal{L}_\omega|_{V^0_p}$. Since \eqref{linealst3} holds, we see that
\begin{equation}\label{bm}
\| \mathcal{L}^{-1}_\omega \| \leq C_{5.1}.
\end{equation}

\textbf{Lemma 5.2}

There is a constant $C_{5.2}$ such that

\begin{equation}\label{cont1}
 \| \mathcal{L}_{\omega_1} - \mathcal{L}_{\omega_2} \|
 \leq C_{5.2} \| \omega_1 - \omega_2\|_{Y_p},
 \end{equation}
 \begin{equation}\label{cont2}
  \| \mathcal{L}^{-1}_{\omega_1} - \mathcal{L}^{-1}_{\omega_2} \|
  \leq C_{5.2} \| \omega_1 - \omega_2\|_{Y_p},
\end{equation}
for all $\omega_1, \omega_2 \in Y^{\frac{\beta_1}{2}, 2 \beta_2}_p$.

\textbf{Proof.} Obviously,
\begin{multline}\label{cont3}
\| \mathcal{L}_{\omega_1} - \mathcal{L}_{\omega_2} \| =
\sup\limits_{\| \boldsymbol u \|_{V_p}=1}
\| \boldsymbol \nabla \cdot \big((\omega_2-\omega_1) \boldsymbol{C}
\boldsymbol\varepsilon (\boldsymbol u) \big) \|_{X_p} \\
\leq \sup\limits_{\| \boldsymbol u \|_{V_p}=1} \Big(
\| \boldsymbol \nabla (\omega_1 - \omega_2) \|_{X_p} \cdot
\| \boldsymbol{C}
\boldsymbol \varepsilon (\boldsymbol u) \|_{C^0}  \\ +
\| \omega_1 - \omega_2 \|_{C^0} \cdot
\| \boldsymbol \nabla \cdot \big( \boldsymbol{C}
\boldsymbol\varepsilon (\boldsymbol u) \big) \|_{X_p} \Big)
\leq C \
\| \omega_1 - \omega_2 \|_{Y_p}.
\end{multline}

Furthermore,
\begin{equation}\label{cont4}
\| \mathcal{L}^{-1}_{\omega_1} - \mathcal{L}^{-1}_{\omega_2} \| \leq
\| \mathcal{L}^{-1}_{\omega_1} \| \cdot
\| \mathcal{L}^{-1}_{\omega_2} \| \cdot
\| \mathcal{L}_{\omega_1} - \mathcal{L}_{\omega_2} \|
\leq C \
\| \omega_1 - \omega_2 \|_{Y_p}.
\end{equation}
%This completes the proof of \emph{Lemma 5.2}
The lemma is proved $\blacksquare$

\textbf{Lemma 5.3}

Let $\boldsymbol{u}^{\ast} \in V_p$, $T>0$, $\boldsymbol{q} \in C^0([0,T],X_p)$,
${\boldsymbol\varepsilon}^{cr} \in C^0([0,T],Y_p^4)$, \\
$\omega \in C^0([0,T],Y_p^{\frac{\beta_1}{2}, 2 \beta_2})$.
Then there exists a uniquely determined mapping \\
$\boldsymbol U= \boldsymbol U \big(\boldsymbol{u}^{\ast},
{\boldsymbol\varepsilon}^{cr}, \omega, \boldsymbol{q} \big)
\in C^0([0,T],V_p^0)$ such that
\begin{equation}\label{prob}
\mathcal{L}_{\omega(t)}(\boldsymbol U(t)) =
-\mathcal{L}_{\omega(t)}(\boldsymbol{u}^{\ast}) - \boldsymbol q (t)
+ \boldsymbol \nabla \cdot \big((1-\omega(t)) \boldsymbol{C}
\boldsymbol\varepsilon^{cr}(t) \big)
\end{equation}
for $t \in [0,T]$. Moreover,
\begin{equation}\label{prob2}
\| \boldsymbol U \big(\boldsymbol{u}^{\ast},
{\boldsymbol\varepsilon}^{cr}, \omega,
\boldsymbol{q} \big) \|_{V_p, \infty} \leq
C_{5.3} \big( \| \boldsymbol{u}^{\ast} \|_{V_p} +
\| {\boldsymbol\varepsilon}^{cr} \|_{Y_p^4, \infty}
+ \| \boldsymbol q \|_{X_p, \infty} \big),
\end{equation}
\begin{multline}\label{prob3}
\| \boldsymbol U \big(\boldsymbol{u}^{\ast},
{{\boldsymbol\varepsilon}^{cr}}^1, \omega^1, \boldsymbol{q}^1 \big) -
\boldsymbol U \big(\boldsymbol{u}^{\ast},
{{\boldsymbol\varepsilon}^{cr}}^2, \omega^2, \boldsymbol{q}^2 \big)
\|_{V_p, \infty} \\ \leq
C_{5.3} \Big( \| \omega^1 - \omega^2 \|_{Y_p, \infty}
\big( \| \boldsymbol{u}^{\ast} \|_{V_p} +
\| {{\boldsymbol\varepsilon}^{cr}}^1 \|_{Y_p^4, \infty}
+ \| \boldsymbol q^1 \|_{X_p, \infty} \big) \\
 + \| {{\boldsymbol\varepsilon}^{cr}}^1
 - {{\boldsymbol\varepsilon}^{cr}}^2\|_{Y_p^4, \infty}
 +  \| \boldsymbol q^1 - \boldsymbol q^2 \|_{X_p, \infty} \Big).
\end{multline}
\textbf{Proof.}
We claim that the mapping $\boldsymbol U \in C^0([0,T],V_p^0)$ is uniquely
defined by \eqref{prob}. Indeed,
%this follows from Lemma 5.1 after some easy computations.
at each instant of time we have by \emph{Lemma 5.1}
\begin{equation}\label{prob5}
\boldsymbol U \big(\boldsymbol{u}^{\ast},
{\boldsymbol\varepsilon}^{cr}, \omega, \boldsymbol{q}
\big) (t) = \mathcal{L}^{-1}_{\omega(t)}
\Big( -\mathcal{L}_{\omega(t)}(\boldsymbol{u}^{\ast}) - \boldsymbol q (t)
+ \boldsymbol \nabla \cdot \big((1-\omega(t)) \boldsymbol{C}
\boldsymbol\varepsilon^{cr}(t) \big)  \Big).
\end{equation}
Estimate \eqref{prob2} follows from \eqref{linealst3}.

Combining \eqref{bm}, \eqref{cont1}, and \eqref{cont2} we note that
\begin{equation}\label{prob6}
\| \mathcal{L}^{-1}_{\omega^1} \mathcal{L}_{\omega^1} \boldsymbol{u}^{\ast}
 - \mathcal{L}^{-1}_{\omega^2} \mathcal{L}_{\omega^2} \boldsymbol{u}^{\ast} \| \leq
C \ \| \boldsymbol{u}^{\ast} \|_{V_p} \| \omega^1 - \omega^2 \|_{Y_p}.
\end{equation}

Note also that
\begin{equation}\label{prob7}
\| \mathcal{L}^{-1}_{\omega^1} \boldsymbol{g}^1 -
\mathcal{L}^{-1}_{\omega^2} \boldsymbol{g}^2 \| \leq
C \| \omega^1 - \omega^1 \|_{Y_p} \|\boldsymbol{g}^1\|_{X_p} +
C \|\boldsymbol{g}^1 - \boldsymbol{g}^2 \|_{X_p},
\ \forall \boldsymbol{g}^1, \boldsymbol{g}^2 \in X_p.
\end{equation}
Substituting $\Big(-\boldsymbol q^i (t)
+ \boldsymbol \nabla \cdot \big((1-\omega^i(t)) \boldsymbol{C}
{\boldsymbol\varepsilon^{cr}}^i(t)\Big)$ for $\boldsymbol g^i$ in \eqref{prob7}
and combining \eqref{prob7} with \eqref{prob6}, we get \eqref{prob3}.
This completes the proof of \emph{Lemma 5.3} $\blacksquare$

\subsection{Evolution of $\omega$ and
$\boldsymbol\varepsilon^{cr}$ with a given $\boldsymbol u$}

%First let us rewrite evolutions \eqref{consteq2}, \eqref{consteq3} in a compact form
%
%\begin{equation}\label{ev1}
%\dot{\boldsymbol\varepsilon}^{cr}(x,t)= \mathcal{R}
%\big( \mathcal{B}(\boldsymbol u, \boldsymbol\varepsilon, ) \big)
%\end{equation}
Let us first analyze evolution operators $\mathcal{R}$, $\mathcal{S}$, which are
defined by \eqref{comact4}, \eqref{comact5}.

\textbf{Lemma 5.4}

There is a constant $C_{5.4}$ with the properties to follow. For
all $i,j \in \{1,2\}$, $k \in \{0,1,...,7\}$,
$\rho^1, \rho^2 \in \mathbb{R}^7$ such that
$\rho^1_7, \rho^2_7 \leq 1-\frac{\beta_1}{2}$
we have the following estimates
\begin{equation}\label{ev1}
|D_k \mathcal{R}_{i,j}(\rho^1)|+|D_k \mathcal{S}(\rho^1)|
\leq C_{5.4} (1+|\rho^1|)^{\max(m,n)},
\end{equation}
\begin{multline}\label{ev2}
|D_k \mathcal{R}_{i,j}(\rho^1) - D_k \mathcal{R}_{i,j}(\rho^2)|+
|D_k \mathcal{S}(\rho^1) - D_k \mathcal{S}(\rho^2)| \\ \leq C_{5.4}
(1+|\rho^1|+|\rho^2|)^{\max(m,n)} |\rho^1-\rho^2|.
\end{multline}
Here $D_k= \frac{\partial}{\partial \rho_k}$ for $k \in \{1,2,...,7\}$ and
$D_0$ is the identity operator.

\textbf{Proof.}
First let us note that for all $q>2$
\begin{equation}\label{vs1}
|P^q (z)| \leq C |z|^q, \quad
|P^q (z) - P^q (z')| \leq C (|z|+|z'|)^{q-1} |z-z'|,
\end{equation}
\begin{equation}\label{vs2}
|\boldsymbol \nabla_z P^q (z)| \leq C |z|^{q-1}, \quad
|\boldsymbol \nabla_z P^q (z) - \boldsymbol \nabla_z P^q (z')|
\leq C (|z|+|z'|)^{q-2} |z-z'|,
\end{equation}
where $P(z)$ is defined by \eqref{comact10}.

We also note that for all $\rho^1, \rho^2 \in \mathbb{R}^7$,
$i,j \in \{1,2\}$, $k \in \{1,2,...,7\}$
\begin{equation}\label{vs3}
|\sigma_{i,j}(\rho)| \leq C |\rho|, \quad
|\sigma_{i,j}(\rho^1) - \sigma_{i,j}(\rho^2)|
\leq C (1+|\rho^1|+|\rho^2|) |\rho^1-\rho^2|,
\end{equation}
\begin{equation}\label{vs4}
|\frac{\partial}{\partial \rho_k} \sigma_{i,j}(\rho)|
\leq C (1+ |\rho|), \quad
|\frac{\partial}{\partial \rho_k} (\sigma_{i,j}(\rho^1) -
\sigma_{i,j}(\rho^2)) | \leq
C |\rho^1-\rho^2|.
\end{equation}
Here $\sigma_{i,j}(\rho)$ is defined by \eqref{comact9}.

The lemma is proved after some simple computations. Let us prove for
example that
\begin{equation}\label{vs5}
|D_k \mathcal{S}(\rho^1) - D_k \mathcal{S}(\rho^2)|
\leq C \ (1+|\rho^1|+|\rho^2|)^m \ |\rho^1-\rho^2|.
\end{equation}
We remark that $\mathcal{S}(\rho)$ has the form
\begin{equation}\label{vs6}
\mathcal{S}(\rho)= P^m(\sigma_{ij}(\rho)) F(\rho_7)
\end{equation}
with $F \in C^{\infty}[0, 1-\frac{\beta_1}{2}]$. Here
$\sigma_{ij} = (\sigma_{1,1}, \sigma_{2,2}, \sigma_{1,2})^T$.
\begin{multline}\label{vs7}
|D_k \mathcal{S}(\rho^1) - D_k \mathcal{S}(\rho^2)| \\
\leq | D_k \big(P^m(\sigma_{ij}(\rho^1)) F(\rho^1_7) -
P^m(\sigma_{ij}(\rho^2)) F(\rho^1_7) \big) | \\ +
| D_k \big(P^m(\sigma_{ij}(\rho^2)) F(\rho^1_7) -
P^m(\sigma_{ij}(\rho^2)) F(\rho^2_7) \big) | \\
\leq C \ | D_k \big(P^m(\sigma_{ij}(\rho^1)) -
P^m(\sigma_{ij}(\rho^2)) \big) |  +
C \ | P^m(\sigma_{ij}(\rho^1)) -
P^m(\sigma_{ij}(\rho^2)) | \\ +
C \ | D_k P^m(\sigma_{ij}(\rho^2)) | \cdot |\rho^1-\rho^2| +
C \ | P^m(\sigma_{ij}(\rho^2)) | \cdot |\rho^1-\rho^2| \\
= \emph{A} + \emph{B} + \emph{C} + \emph{D}.
\end{multline}
We abbreviate
\begin{equation}\label{vs8}
D_k \sigma (\rho):= D_k (\sigma_{1,1} (\rho), \sigma_{2,2}
(\rho), \sigma_{1,2} (\rho))^T.
\end{equation}
Further,
\begin{multline}\label{vs9}
\emph{A} = C \ | \nabla_z P^m (\sigma_{ij}(\rho^1)) D_k \sigma (\rho^1) -
\nabla_z P^m (\sigma_{ij}(\rho^2)) D_k \sigma (\rho^2)| \\
\leq C \ | \nabla_z \big( P^m (\sigma_{ij}(\rho^1))
- P^m (\sigma_{ij}(\rho^2)) \big)| \cdot
| D_k \sigma (\rho^1)| \\
 + C \ | \nabla_z P^m (\sigma_{ij}(\rho^2)) |
 \cdot |D_k \big(\sigma (\rho^1) - \sigma (\rho^2)\big)| \\
\leq C \ \big(|\sigma_{ij}(\rho^1)|+|\sigma_{ij}(\rho^2)| \big)^{m-2}
\cdot |\sigma_{ij}(\rho^1) - \sigma_{ij}(\rho^2)|
\cdot | D_k \sigma (\rho^1)| \\
+ C \ |\sigma_{ij}(\rho^2)|^{m-1}
\cdot |D_k \big(\sigma (\rho^1) - \sigma (\rho^2)\big)| \\
\leq C \ (1+|\rho^1|+|\rho^2|)^m \ |\rho^1-\rho^2|.
\end{multline}

In the same way we obtain
\begin{equation}\label{vs10}
\max(\emph{B}, \emph{C}, \emph{D}) \leq C \ (1+|\rho^1|+|\rho^2|)^m \ |\rho^1-\rho^2|.
\end{equation}
Combining this with \eqref{vs9} we get \eqref{vs5}.
The lemma is proved $\blacksquare$

\textbf{Lemma 5.5}

There is a constant $C_{5.5}$ with the following properties.
For all $M, T > 0$, $\boldsymbol u^1, \boldsymbol u^2 \in C^0([0,T],V_p)$
with $\| u^l \|_{V_p, \infty} \leq M$ for $l \in \{1,2\}$;
${{\boldsymbol\varepsilon}^{cr}}^1, {{\boldsymbol\varepsilon}^{cr}}^2 \in C^0([0,T],Y^4_p)$;
${\omega}^1, {\omega}^2 \in C^0([0,T],Y_p)$ such that

\begin{equation}\label{vv1}
\omega(x,t) \leq 1- \frac{\beta_1}{2} \quad \forall \ (x,t) \in \Omega \times [0,T], l \in \{1,2\}
\end{equation}
\begin{equation}\label{vv2}
\|{{\boldsymbol\varepsilon}^{cr}}^l(t) -
{{\boldsymbol\varepsilon}^{cr}}^l(0) \|_{Y^4_p}
+ \|{\omega}^l(t) - {\omega}^l(0) \|_{Y_p}
\leq 1 \quad \forall \ t \in [0,T], l \in \{1,2\}.
\end{equation}

We abbreviate (recall \eqref{comact22})
\begin{equation}\label{vv3}
\rho^l:=\rho( \boldsymbol u^l,
{{\boldsymbol\varepsilon}^{cr}}^l, {\omega}^l).
\end{equation}

Then
\begin{equation}\label{vv4}
\mathcal{R}(\rho^l)(t) \in Y^4_p, \quad
\mathcal{S}(\rho^l)(t) \in Y_p,  \quad \forall \ t \in [0,T], l \in \{1,2\},
\end{equation}
\begin{equation}\label{vv5}
\| \mathcal{R}(\rho^l) \|_{Y^4_p, \infty} +
\| \mathcal{S}(\rho^l) \|_{Y_p, \infty}
 \leq C_{5.5} \big( M +
\|{{\boldsymbol\varepsilon}^{cr}}^l(0) \|_{Y^4_p} + \| {\omega}^l(0) \|_{Y_p} +1
  \big)^{\max(m,n)+1}
\end{equation}
for $l \in \{1,2\}$, and
\begin{multline}\label{vv6}
\| \mathcal{R}(\rho^1) - \mathcal{R}(\rho^2) \|_{Y^4_p, \infty} +
\| \mathcal{S}(\rho^1) - \mathcal{S}(\rho^2)\|_{Y_p, \infty} \\
 \leq C_{5.5} \Big( M + \sum\limits_{l=1}^2 \big(
\|{{\boldsymbol\varepsilon}^{cr}}^l(0) \|_{Y^4_p}
+ \| {\omega}^l(0) \|_{Y_p} \big) +1
  \Big)^{\max(m,n)+1} \\
\cdot \Big(
\| \boldsymbol{u}^1 - \boldsymbol{u}^2 \|_{V_p, \infty} +
\| {{\boldsymbol\varepsilon}^{cr}}^1
 - {{\boldsymbol\varepsilon}^{cr}}^2\|_{Y_p^4, \infty}
 + \| \omega^1 - \omega^2 \|_{Y_p, \infty} \Big).
\end{multline}

\textbf{Proof.}
This lemma is proved by \emph{Corollary 3.1} and \emph{Lemma 5.4}.
Let us estimate, for example, the
 value $\| K \|_{L_p(\Omega)}(t)$, where
\begin{equation}\label{vv7}
K(x,t):= \Big| \frac{\partial}{\partial x_r} \big( \mathcal{R}_{i,j}(\rho^1) -
\mathcal{R}_{i,j}(\rho^2) \big) (x,t) \Big|,
\end{equation}
with $r, i, j \in \{1,2\}, (x,t) \in \Omega \times [0,T]$.
We obtain
\begin{multline}\label{vv8}
K(x,t) \leq \sum\limits_{k=1}^7 \Big|
D_k \mathcal{R}_{i,j}\big(\rho^1_k(x,t)\big)
\frac{\partial \rho^1_k(x,t)}{\partial x_r}  -
D_k \mathcal{R}_{i,j}\big(\rho^2_k(x,t)\big)
\frac{\partial \rho^2_k(x,t)}{\partial x_r}
\Big| \\
\leq
\sum\limits_{k=1}^7 \Big| D_k \mathcal{R}_{i,j}
\big(\rho^1(x,t)\big) - D_k \mathcal{R}_{i,j}\big(\rho^2(x,t)\big)\Big|
\cdot \Big| \frac{\partial \rho^1_k(x,t)}{\partial x_r} \Big| \\
+ \sum\limits_{k=1}^7 \Big| D_k \mathcal{R}_{i,j}\big(\rho^2(x,t)\big) \Big| \cdot
\Big|  \frac{\partial \rho^1_k(x,t)}{\partial x_r} -
\frac{\partial \rho^2_k(x,t)}{\partial x_r}\Big| \\
\leq \sum\limits_{k=1}^7 A_k \cdot B_k + C_k \cdot D_k.
\end{multline}
But by \emph{Lemma 5.4} and by \emph{Corollary 3.1} we get
\begin{multline}\label{vv9}
A_k \leq C \ \big(1+|\rho^1(x,t)|
+|\rho^2(x,t)|\big)^{\max(m,n)} \big|\rho^1(x,t)-\rho^2(x,t)\big| \\
\leq C \ \big(1+ \|\rho^1\|_{Y^7_p, \infty}
+\|\rho^2(x,t)\|_{Y^7_p, \infty } \big)^{\max(m,n)}
\|\rho^1-\rho^2\|_{Y^7_p, \infty }.
\end{multline}
We get by the same argument
\begin{equation}\label{vv10}
C_k \leq C \ \big| D_k \mathcal{R}_{i,j}\big(\rho^2(x,t)\big) \big| \leq C \
\big(1+\|\rho^2\|_{Y^7_p, \infty} \big)^{\max(m,n)}.
\end{equation}

Evidently,
\begin{equation}\label{vv11}
\| B_k \|_{L_p} = \| \frac{\partial \rho^1_k(x,t)}
{\partial x_r} \|_{L_p} \leq \|\rho^1\|_{Y^7_p, \infty},
\end{equation}
\begin{equation}\label{vv12}
\| D_k \|_{L_p} = \| \frac{\partial \rho^1_k(x,t)}{\partial x_r} -
\frac{\partial \rho^2_k(x,t)}{\partial x_r}
 \|_{L_p} \leq \|\rho^1 - \rho^2\|_{Y^7_p, \infty}.
\end{equation}
Hence
\begin{multline}\label{vv13}
\| K \|_{L_p(\Omega)}(t) \leq   \sum\limits_{k=1}^7 \|
A_k \cdot B_k + C_k \cdot D_k \|_{L_p} \\ \leq
\sum\limits_{k=1}^7 \| A_k \|_{C^0} \cdot \|
B_k \|_{L_p} +  \| C_k \|_{C^0} \cdot \| D_k \|_{L_p}
\\ \leq
 C \
\big(1+\|\rho^1\|_{Y^7_p, \infty}
+\|\rho^2\|_{Y^7_p, \infty} \big)^{\max(m,n)+1}
\cdot \|\rho^1 - \rho^2\|_{Y^7_p, \infty}.
\end{multline}
It remains to check that
\begin{equation}\label{vv14}
\|\rho^1 - \rho^2\|_{Y_p^7, \infty} \leq C \ \Big(
\| \boldsymbol{u}^1 - \boldsymbol{u}^2 \|_{V_p, \infty} +
\| {{\boldsymbol\varepsilon}^{cr}}^1
 - {{\boldsymbol\varepsilon}^{cr}}^2\|_{Y_p^4, \infty}
 + \| \omega^1 - \omega^2 \|_{Y_p, \infty} \Big),
\end{equation}
and
\begin{equation}\label{vv15}
\big(1+\|\rho^1\|_{Y^7_p, \infty}+\|\rho^2\|_{Y^7_p, \infty} \big) \leq
C \ \Big( M + \sum\limits_{l=1}^2 \big(
\|{{\boldsymbol\varepsilon}^{cr}}^l(0) \|_{Y^4_p}
+ \| {\omega}^l(0) \|_{Y_p} \big) +1
  \Big).
\end{equation}
This concludes the proof of \emph{Lemma 5.5}$\blacksquare$

We now prove the existence of the solution $(\boldsymbol\varepsilon^{cr}, \omega) \in
C^0([0,T],Y^4_p \times Y_p)$ if the displacement field $\boldsymbol u \in
C^0([0,T],V_p)$ is given.

\textbf{Lemma 5.6}

Let $M > 0$, $\boldsymbol\varepsilon^{cr}_0 \in Y^4_p$,
and $\omega_0 \in Y^{\beta_1, \beta_2}_p$.
Put
\begin{equation}\label{L1}
T_0:=\Big[ C_{5.5} \big( M + 2 \|{\boldsymbol\varepsilon}^{cr}_0 \|_{Y^4_p}
+ 2 \| \omega_0 \|_{Y_p} + 1\big)^{\max(m,n)+1} 2 \frac{1+C_I}{\min(\beta_1, 2 \beta_2)} \Big]^{-1}
\end{equation}
with $C_{5.5}$ from \emph{Lemma 5.5} and $C_I$
from \emph{Corollary 3.1}. Let $T' \in (0,T_0]$
and $\boldsymbol u \in C^0([0,T'],V_p)$, such
that $\| \boldsymbol u \|_{V_p, \infty} \leq M$.
Then there exists a unique mapping
$(\boldsymbol\varepsilon^{cr}, \omega)
=\big(\boldsymbol\varepsilon^{cr}, \omega\big)
(\boldsymbol u, {\boldsymbol\varepsilon}^{cr}_0, \omega_0)
\in C^0([0,T'],Y^4_p \times Y_p)$
such that
\begin{equation}\label{L2}
(\boldsymbol\varepsilon^{cr}, \omega)(t)
=(\boldsymbol\varepsilon^{cr}_0, \omega_0) +
\int\limits_0^t \big( \mathcal{R}(\rho (\boldsymbol u, {\boldsymbol\varepsilon}^{cr}, \omega)
), \mathcal{S}(\rho (\boldsymbol u, {\boldsymbol\varepsilon}^{cr}, \omega) )\big)(s)ds,
\end{equation}
\begin{equation}\label{L3}
\|{\boldsymbol\varepsilon}^{cr}(t) - {\boldsymbol\varepsilon}^{cr}(0) \|_{Y^4_p}
+ \|\omega(t) - \omega(0) \|_{Y_p}
\leq \frac{\min(\beta_1, 2 \beta_2)}{2 (1+C_I)} \quad \forall \ t \in [0,T'].
\end{equation}
Moreover
\begin{equation}\label{L4}
\omega(t) \in Y^{\frac{\beta_1}{2}, 2 \beta_2}_p \quad \forall \ t \in [0,T'],
\end{equation}
\begin{equation}\label{L5}
(\boldsymbol\varepsilon^{cr}, \omega) \in C^1([0,T'],Y^4_p \times Y_p),
\end{equation}
\begin{equation}\label{L5}
(\dot{\boldsymbol\varepsilon^{cr}}, \dot{\omega})(t) =
\big( \mathcal{R}(\rho (\boldsymbol u, {\boldsymbol\varepsilon}^{cr}, \omega)),
\mathcal{S}(\rho (\boldsymbol u, {\boldsymbol\varepsilon}^{cr}, \omega) )\big)(t) \quad \forall \ t \in (0,T').
\end{equation}

\textbf{Proof.} As it is done in \cite{Deuring1} for partly
coupled damage model, we adapt
the standard proof of the existence of
solutions to ordinary differential equations in Banach
spaces. We define the closed subset of $C^0([0,T'],Y^4_p \times Y_p)$ by
\begin{multline}\label{L6}
\mathcal{M}:=\Big\{ (\boldsymbol\varepsilon^{cr}, \omega)
\in C^0([0,T'],Y^4_p \times Y_p):
(\boldsymbol\varepsilon^{cr}, \omega) (0)
= (\boldsymbol\varepsilon^{cr}_0, \omega_0), \\
\|{\boldsymbol\varepsilon}^{cr}(t) - {\boldsymbol\varepsilon}^{cr}(0) \|_{Y^4_p}
+ \|\omega(t) - \omega(0) \|_{Y_p}
\leq \frac{\min(\beta_1, 2 \beta_2)}{2 (1+C_I)}
\Big\}.
\end{multline}
The application of \emph{Corollary 3.1} yields
for every $(\boldsymbol\varepsilon^{cr}, \omega) \in \mathcal{M}$
\begin{equation}\label{L7}
 \omega(t) \in Y^{\frac{\beta_1}{2}, 2 \beta_2}_p \quad \forall \ t \in [0,T'].
\end{equation}
Note that for all $t \in [0,T'],
(\boldsymbol\varepsilon^{cr}, \omega) \in \mathcal{M}$
\begin{equation}\label{L8}
\|{\boldsymbol\varepsilon}^{cr}(t) - {\boldsymbol\varepsilon}^{cr}(0) \|_{Y^4_p}
+ \|\omega(t) - \omega(0) \|_{Y_p}
\leq \frac{\min(\beta_1, 2 \beta_2)}{2 (1+C_I)} \leq 1.
\end{equation}
Hence, by \emph{Lemma 5.5}, we obtain for $t \leq t',
(\boldsymbol\varepsilon^{cr}, \omega) \in \mathcal{M}$
\begin{multline}\label{L9}
\int\limits_t^{t'} \big\|
\big( \mathcal{R}(\rho (\boldsymbol u, {\boldsymbol\varepsilon}^{cr}, \omega)
), \mathcal{S}(\rho (\boldsymbol u, {\boldsymbol\varepsilon}^{cr}, \omega) )\big) (s) \big\|_{Y^4_p \times Y_p} ds
\\
\leq C_{5.5} \big( M +
\|{{\boldsymbol\varepsilon}^{cr}}^l(0) \|_{Y^4_p} + \| {\omega}^l(0) \|_{Y_p} +1
  \big)^{\max(m,n)+1} (t-t') \\
\leq \frac{\min(\beta_1, 2 \beta_2)}{2 (1+C_I)}.
\end{multline}

Let the mapping $\mathcal{T}: \mathcal{M} \rightarrow C^0([0,T'],Y^4_p \times Y_p)$
be given by
\begin{multline}\label{L10}
\mathcal{T}(\boldsymbol\varepsilon^{cr}, \omega)(t) \\ =
(\boldsymbol\varepsilon^{cr}_0, \omega_0)+ \int\limits_0^t
\big( \mathcal{R}(\rho (\boldsymbol u, {\boldsymbol\varepsilon}^{cr}, \omega)
), \mathcal{S}(\rho (\boldsymbol u, {\boldsymbol\varepsilon}^{cr}, \omega) )\big)(s) ds
 \quad \forall \ t \in [0,T'].
\end{multline}
Accordingly to \eqref{L9}, this mapping is well defined. Taking into
account \eqref{L9}, we obtain
\begin{equation}\label{L11}
\| \mathcal{T}(\boldsymbol\varepsilon^{cr}, \omega)(t) -
(\boldsymbol\varepsilon^{cr}_0, \omega_0) \|_{Y^4_p \times Y_p}
\leq \frac{\min(\beta_1, 2 \beta_2)}{2 (1+C_I)} \quad \forall \ t \in [0,T'].
\end{equation}
Therefore, $\mathcal{T}(\mathcal{M}) \subset \mathcal{M}$.
By \emph{Lemma 5.5}, it follows that $\mathcal{T}$ is
a contraction with respect to the norm of the
space $C^0([0,T'],Y^4_p \times Y_p)$. Indeed, for every instant
of time $t$ we have
\begin{multline}\label{L12}
\| \mathcal{T}({\boldsymbol\varepsilon^{cr}}^1, \omega^1)(t) -
\mathcal{T}({\boldsymbol\varepsilon^{cr}}^2,
\omega^2)(t) \|_{Y^4_p \times Y_p}
\\ \leq C_{5.5}
\Big( M + \sum\limits_{l=1}^2 \big(
\|{{\boldsymbol\varepsilon}^{cr}}^l(0) \|_{Y^4_p}
+ \| {\omega}^l(0) \|_{Y_p} \big) +1
  \Big)^{\max(m,n)+1} \\
\cdot \Big(
\| {{\boldsymbol\varepsilon}^{cr}}^1
 - {{\boldsymbol\varepsilon}^{cr}}^2\|_{Y_p^4, \infty}
 + \| \omega^1 - \omega^2 \|_{Y_p, \infty} \Big) T'
\leq \\ \frac{1}{4}
\Big(
\| {{\boldsymbol\varepsilon}^{cr}}^1
 - {{\boldsymbol\varepsilon}^{cr}}^2\|_{Y_p^4, \infty}
 + \| \omega^1 - \omega^2 \|_{Y_p, \infty} \Big).
\end{multline}
It follows from \emph{Banach's fixed point
theorem} (see, for example, \cite{Evans}) that
there exists a unique $(\boldsymbol\varepsilon^{cr}, \omega) \in \mathcal{M}$
such that $\mathcal{T}(\boldsymbol\varepsilon^{cr}, \omega)=(\boldsymbol\varepsilon^{cr}, \omega)$.
Thus, we have proved that the pair
$(\boldsymbol\varepsilon^{cr}, \omega) \in C^0([0,T'],Y^4_p \times Y_p)$
is uniquely defined by \eqref{L2}, \eqref{L3}.

To conclude the proof it remains to note that the mapping
$t \mapsto
\big( \mathcal{R}(\rho (\boldsymbol u, {\boldsymbol\varepsilon}^{cr}, \omega)
), \mathcal{S}(\rho (\boldsymbol u, {\boldsymbol\varepsilon}^{cr},
\omega) )\big)(t)$ is continuous
from $[0,T']$ to $Y^4_p \times Y_p$. Since
\eqref{L4} holds, we may use \emph{Lemma 5.4} to prove that
\begin{multline}\label{L13}
 \|\big( \mathcal{R}(\rho (\boldsymbol u, {\boldsymbol\varepsilon}^{cr}, \omega)
), \mathcal{S}(\rho (\boldsymbol u, {\boldsymbol\varepsilon}^{cr}, \omega) )\big)(t^1) \\ -
 \big( \mathcal{R}(\rho (\boldsymbol u, {\boldsymbol\varepsilon}^{cr}, \omega)
), \mathcal{S}(\rho (\boldsymbol u, {\boldsymbol\varepsilon}^{cr}, \omega) )\big)(t^2)\|_{Y^4_p \times Y_p}
\rightarrow 0,
\quad \text{as} \ t^1 \rightarrow t^2.
%\leq C \ \| \rho (\boldsymbol u, {\boldsymbol\varepsilon}^{cr}, \omega) (t^1)-
%\rho (\boldsymbol u, {\boldsymbol\varepsilon}^{cr}, \omega) (t^2)\|_{Y^7_p}
\end{multline}
\emph{Lemma 5.6} is proved $\blacksquare$

Now we need to estimate the difference
between two solutions of \eqref{L2}, \eqref{L3}.
Let us abbreviate
\begin{equation}\label{L14}
\hat{C} = C_{5.5} \big( M + 2 \|{\boldsymbol\varepsilon}^{cr}_0 \|_{Y^4_p}
+ 2 \| \omega_0 \|_{Y_p} + 1\big)^{\max(m,n)+1}
\end{equation}
with $C_{5.5}$ from \emph{Lemma 5.5}
and $C_I$ from \emph{Corollary 3.1}.

\textbf{Lemma 5.7}

Let $M > 0$, $K \geq 1$, $\boldsymbol\varepsilon^{cr}_0
\in Y^4_p$, and $\omega_0 \in Y^{\beta_1, \beta_2}_p$.
Put
\begin{equation}\label{L15}
T_1:=\Big[ \hat{C} \cdot  2 \max(K, \frac{1+C_I}{\min(\beta_1, 2 \beta_2)}) \Big]^{-1}
\end{equation}
where $\hat{C}$ is given by \eqref{L14}.
Let $T' \in (0,T_1]$,
and $\boldsymbol u^1, \boldsymbol u^2 \in C^0([0,T'],V_p)$,
such that $\| \boldsymbol u^l \|_{V_p, \infty} \leq M$
for $l \in \{1,2\}$. Assume that
\begin{equation}\label{L16}
({\boldsymbol\varepsilon^{cr}}^l, {\omega}^l)=
\big(\boldsymbol\varepsilon^{cr}, \omega\big)
(\boldsymbol u^l, {\boldsymbol\varepsilon}^{cr}_0, \omega_0)
\in C^0([0,T'],Y^4_p \times Y_p), l \in \{1,2\}
\end{equation}
are defined by \eqref{L2}, \eqref{L3}. Then
\begin{equation}\label{L17}
\| {{\boldsymbol\varepsilon}^{cr}}^1
 - {{\boldsymbol\varepsilon}^{cr}}^2\|_{Y_p^4, \infty}
 + \| \omega^1 - \omega^2 \|_{Y_p, \infty}  \leq \frac{1}{K} \|
 \boldsymbol u^1 - \boldsymbol u^2 \|_{V_p, \infty}.
\end{equation}

\textbf{Proof.}
By \emph{Lemma 5.5} we have
\begin{multline}\label{L18}
\| \mathcal{R}(\rho^1) - \mathcal{R}(\rho^2) \|_{Y^4_p, \infty} +
\| \mathcal{S}(\rho^1) - \mathcal{S}(\rho^2)\|_{Y_p, \infty} \\
 \leq \hat{C}
\cdot \Big(
\| \boldsymbol{u}^1 - \boldsymbol{u}^2 \|_{V_p, \infty} +
\| {{\boldsymbol\varepsilon}^{cr}}^1
 - {{\boldsymbol\varepsilon}^{cr}}^2\|_{Y_p^4, \infty}
 + \| \omega^1 - \omega^2 \|_{Y_p, \infty} \Big).
 \end{multline}
Therefore, since \eqref{L2} holds, we obtain
\begin{multline}\label{L20}
\| {{\boldsymbol\varepsilon}^{cr}}^1
 - {{\boldsymbol\varepsilon}^{cr}}^2\|_{Y_p^4, \infty}
 + \| \omega^1 - \omega^2 \|_{Y_p, \infty} \\
 \leq \hat{C} T'
\Big(
\| \boldsymbol{u}^1 - \boldsymbol{u}^2 \|_{V_p, \infty} +
\| {{\boldsymbol\varepsilon}^{cr}}^1
 - {{\boldsymbol\varepsilon}^{cr}}^2\|_{Y_p^4, \infty}
 + \| \omega^1 - \omega^2 \|_{Y_p, \infty} \Big) \\
\leq
 \frac{1}{2 K} \| \boldsymbol u^1 -
 \boldsymbol u^2 \|_{V_p, \infty} + \frac{1}{2}
 \Big(
\| {{\boldsymbol\varepsilon}^{cr}}^1
 - {{\boldsymbol\varepsilon}^{cr}}^2\|_{Y_p^4, \infty}
 + \| \omega^1 - \omega^2 \|_{Y_p, \infty} \Big).
\end{multline}
Now inequality \eqref{L17} follows from
\eqref{L20}. \emph{Lemma 5.7} is proved $\blacksquare$

\subsection{Equilibrium equations coupled with evolution law}

In this subsection we solve problem \eqref{result2},
\eqref{result3} and prove \emph{Theorem 4.1}.

\textbf{Proof of Theorem 4.1.}
We choose $T_1$ as in \emph{Lemma 5.7} where we put
\begin{equation}\label{M1}
K:=
2 \cdot C_{5.3} \big( \| \boldsymbol{u}^{\ast} \|_{V_p} +
\| {\boldsymbol\varepsilon}^{cr}_0 \|_{V_p^4}
+ \| \boldsymbol q \|_{X_p, \infty} + 1  \big),
\end{equation}
\begin{equation}\label{MM1}
 \ M:= C_{5.3} \big( \| \boldsymbol{u}^{\ast} \|_{V_p} +
\| {\boldsymbol\varepsilon}^{cr}_0 \|_{V_p^4}
+ \| \boldsymbol q \|_{X_p, \infty} +1  \big) +
\| \boldsymbol{u}^{\ast} \|_{V_p}
\end{equation}
with $C_{5.3}$ from \emph{Lemma 5.3}. In order to use
\emph{Banach's fixed point theorem}
we define the closed subset of $C^0([0,T'],V^0_p)$ by
\begin{equation}\label{M2}
\mathcal{M}:=\Big\{ \boldsymbol u  \in C^0([0,T'],V^0_p):
\| \boldsymbol u + \boldsymbol{u}^{\ast} \|_{V_p, \infty} \leq M
\Big\}.
\end{equation}
Let the mapping $\mathcal{T}: \mathcal{M} \rightarrow C^0([0,T'],V^0_p)$
be given by
\begin{equation}\label{M3}
\mathcal{T}(\boldsymbol u)= \boldsymbol U \big(\boldsymbol{u}^{\ast},
{\boldsymbol\varepsilon}^{cr}(\boldsymbol u +
\boldsymbol{u}^{\ast}, {\boldsymbol\varepsilon}^{cr}_0, \omega_0),
\omega(\boldsymbol u + \boldsymbol{u}^{\ast},
{\boldsymbol\varepsilon}^{cr}_0, \omega_0), \boldsymbol{q} \big),
\end{equation}
where $\big({\boldsymbol\varepsilon}^{cr},
\omega\big)(\boldsymbol u, {\boldsymbol\varepsilon}^{cr}_0, \omega_0)$
is defined by \eqref{L2} and $\boldsymbol U \big(\boldsymbol{u}^{\ast},
{\boldsymbol\varepsilon}^{cr}, \omega, \boldsymbol{q} \big)$ is introduced in \emph{Lemma 5.3}.

Let us show that $\mathcal{T} (\mathcal{M}) \subset C^0([0,T'],V^0_p)$. In fact,
accordingly to \emph{Lemma 5.6}, for all $\boldsymbol u \in \mathcal{M}$ we have
$\big({\boldsymbol\varepsilon}^{cr}, \omega\big)
(\boldsymbol u + \boldsymbol{u}^{\ast}, {\boldsymbol\varepsilon}^{cr}_0, \omega_0)
\in C^0([0,T'],Y^4_p \times Y_p)$. Therefore, the assertion follows from \emph{Lemma 5.3}
and the mapping $\mathcal{T}$ is well defined.

Now let us show that $\mathcal{T} (\mathcal{M}) \subset \mathcal{M}$.
Using \eqref{prob2} and \eqref{L3} we obtain for $\boldsymbol u \in \mathcal{M}$
\begin{multline}\label{M4}
\| \mathcal{T}(\boldsymbol u) + \boldsymbol{u}^{\ast} \|_{V_p, \infty}
\leq \| \mathcal{T}(\boldsymbol u) \|_{V_p, \infty} + \| \boldsymbol{u}^{\ast} \|_{V_p}
 \\ \leq
C_{5.3} \big( \| \boldsymbol{u}^{\ast} \|_{V_p} +
\| {\boldsymbol\varepsilon}^{cr}(\boldsymbol{u} + \boldsymbol{u}^{\ast},
{\boldsymbol\varepsilon}^{cr}_0, \omega_0) \|_{Y_p^4, \infty}
+ \| \boldsymbol q \|_{X_p, \infty} \big) + \| \boldsymbol{u}^{\ast} \|_{V_p}
\leq M.
\end{multline}
Let us prove that $\mathcal{T}$ is a contraction.
Taking into account \eqref{prob3}, \eqref{L17}, and the choice of $K$, we obtain
\begin{multline}\label{M5}
\| \mathcal{T}(\boldsymbol u^1) - \mathcal{T}(\boldsymbol u^2) \|_{V_p, \infty} \leq \\
C_{5.3} \Big( \| \omega^1 - \omega^2 \|_{Y_p, \infty}
\big( \| \boldsymbol{u}^{\ast} \|_{V_p} +
\| {\boldsymbol\varepsilon}^{cr}_0 \|_{Y_p^4}
+ \| \boldsymbol q \|_{X_p, \infty} +1 \big) \\
 + \| {{\boldsymbol\varepsilon}^{cr}}^1
 - {{\boldsymbol\varepsilon}^{cr}}^2\|_{Y_p^4, \infty} \Big) \\
\leq \frac{C_{5.3}}{K} \big( \| \boldsymbol{u}^{\ast} \|_{V_p} +
\| {\boldsymbol\varepsilon}^{cr}_0 \|_{Y_p^4}
+ \| \boldsymbol q \|_{X_p, \infty} + 1 \big)
\| \boldsymbol u^1 - \boldsymbol u^2 \|_{V_p, \infty} \\
\leq \frac{1}{2} \| \boldsymbol u^1 - \boldsymbol u^2 \|_{V_p, \infty}.
\end{multline}
From \emph{Banach's fixed point theorem} it
follows that there is a uniquely determined
mapping $\boldsymbol u \in \mathcal{M}$,
such that $\mathcal{T} (\boldsymbol u) = \boldsymbol u$.
To obtain the mapping $(\boldsymbol u, \boldsymbol{\varepsilon}^{cr}, \omega) \in
C^0([0,T'],V^0_p \times Y^4_p \times Y_p)$
that satisfy \eqref{result2} and \eqref{result3}, we put (see \emph{Lemma 5.6})
\begin{equation}\label{M6}
\boldsymbol{\varepsilon}^{cr}:= \boldsymbol{\varepsilon}^{cr} (\boldsymbol u + \boldsymbol{u}^{\ast},
{\boldsymbol\varepsilon}^{cr}_0, \omega_0), \quad
\omega:= \omega (\boldsymbol u + \boldsymbol{u}^{\ast},
{\boldsymbol\varepsilon}^{cr}_0, \omega_0).
\end{equation}
The existence of the solution to
\eqref{result2}, \eqref{result3} is proved. Although
the uniqueness in $\mathcal{M}$ is guarantied
by \emph{Banach's fixed point theorem},
 it remains to check that the solution is
 uniquely determined by \eqref{result2}, \eqref{result3}.
Assume the converse, then there are two
different solutions of \eqref{result2}, \eqref{result3}
\begin{equation}\label{M7}
(\boldsymbol u^l, {\boldsymbol{\varepsilon}^{cr}}^l, {\omega}^l) \in
C^0([0,T'],V^0_p \times Y^4_p \times Y_p), \quad l \in \{1,2\}.
\end{equation}
Put
\begin{equation}\label{M8}
t_0:= \max\{\hat{t} \in [0,T']:
(\boldsymbol u^1, {\boldsymbol{\varepsilon}^{cr}}^1, {\omega}^1)(t)=
(\boldsymbol u^2, {\boldsymbol{\varepsilon}^{cr}}^2, {\omega}^2)(t)
\quad \text{for} \ t \in [0,\hat{t}]\}.
\end{equation}
Hence,
\begin{equation}\label{M9}
(\boldsymbol u^1, {\boldsymbol{\varepsilon}^{cr}}^1, {\omega}^1)(t)=
(\boldsymbol u^2, {\boldsymbol{\varepsilon}^{cr}}^2, {\omega}^2)(t)
\quad \text{for} \ t \in [0,t_0].
\end{equation}
Moreover, $t_0< T'$ and for every $\check{t} \in (t_0,T']$ there
exists $\tilde{t} \in (t_0,\check{t}]$, such that
\begin{equation}\label{M10}
(\boldsymbol u^1, {\boldsymbol{\varepsilon}^{cr}}^1, {\omega}^1)(\tilde{t}) \neq
(\boldsymbol u^2, {\boldsymbol{\varepsilon}^{cr}}^2, {\omega}^2)(\tilde{t}).
\end{equation}

Arguing as above, we prove the uniqueness of the solution \\
$(\boldsymbol u, \boldsymbol{\varepsilon}^{cr}, \omega) \in
C^0([t_0,{T^{\text{new}}}'],V^0_p \times Y^4_p \times Y_p)$ to the new problem
\begin{equation}\label{New1}
\boldsymbol \nabla \cdot \big((1-\omega) \boldsymbol{C}
(\boldsymbol\varepsilon (\boldsymbol u + \boldsymbol u^{\ast})
- {\boldsymbol\varepsilon}^{cr})\big) = - \boldsymbol q(t)
\quad \forall \ t \in [t_0,{T^{\text{new}}}'],
\end{equation}
\begin{multline}\label{New2}
(\boldsymbol{\varepsilon}^{cr}, \omega)(t)=
({\boldsymbol{\varepsilon}^{cr}}^1(t_0), \omega^1(t_0)) \\ +
\int\limits_{t_0}^t \big(\mathcal{R}(\rho
( \boldsymbol u + \boldsymbol u^{\ast}, {\boldsymbol\varepsilon}^{cr}, \omega),
\mathcal{S}(\rho
( \boldsymbol u + \boldsymbol u^{\ast}, {\boldsymbol\varepsilon}^{cr}, \omega)\big)(s)ds
\quad \forall \ t  \in [t_0,{T^{\text{new}}}']
\end{multline}
\begin{equation}\label{New3}
\| \boldsymbol u (t) + \boldsymbol{u}^{\ast} \|_{V_p} \leq M^{\text{new}}
 \quad \forall \ t  \in [t_0,{T^{\text{new}}}'],
\end{equation}
with some new parameters ${T^{\text{new}}}', M^{\text{new}}$.
The reader will easily prove that there
exists $\check{t} \in (t_0,\min({T^{\text{new}}}',T')]$,
such that
\begin{equation}\label{New4}
\| \boldsymbol u^l (t) + \boldsymbol{u}^{\ast} \|_{V_p} \leq M^{\text{new}}
 \quad \forall \ t  \in [t_0,\check{t}], \ l \in \{1,2\}.
\end{equation}
That means, that
\begin{equation}\label{New5}
(\boldsymbol u^1, {\boldsymbol{\varepsilon}^{cr}}^1, {\omega}^1) (t)=
(\boldsymbol u^2, {\boldsymbol{\varepsilon}^{cr}}^2, {\omega}^2) (t)
 \quad \forall \ t  \in [t_0,\check{t}].
\end{equation}
This contradiction proves the theorem $\blacksquare$
% To this end we consider the second solution of \eqref{result2}, \eqref{result3}.

\section{Conclusions}

The creep damage problem is formulated in a well-posed manner.
\emph{Theorem 4.1} states that a
unique smooth solution to the Kachanov-Rabotnov
problem exists in a certain time interval $[0, T_1]$.
The corresponding function spaces $X_p$, $Y_p$, and $V_p$ reflect
the essence of the system of equations
and can be used for a proper mathematical analysis of the problem.
Particulary, clear definitions of terms "stable", "unstable", and
"convergency" can be given.

It is shown that the requirements of the existence theorem (\emph{Theorem 4.1})
have a physical meaning and the violation of these requirements
directly affects the lifetime estimate.

If we do not impose any
restrictions on the gradient of initial damage
(such situation corresponds to $\beta_2=\infty$), then
the lifetime $t^{\ast}$ of the structure can be arbitrary small even if
$\min(1-\omega_0) \geq \beta_1 > 0$.

%This process
%is described by creep-damage solution
%shortly before rupture event as $t \rightarrow t^{\ast}$.
This damage localization effect is characterized
at each instant of time by the quantity
\begin{equation}\label{Damloc}
\Lambda(t) = \frac{\| \nabla \omega \|_{L_p}}{\min(1-\omega)}(t).
\end{equation}

The value $\Lambda(t)$ controls the remaining life of the structure $t_{rest}:=t^{\ast}-t$.
\begin{equation}\label{Damloc2}
t_{rest} \rightarrow 0, \quad \text{as} \quad \Lambda \rightarrow \infty .
\end{equation}
Thus, the estimation of $\Lambda$ gives an answer to the question when the damage
becomes critical. This measure of damage localization can be adopted to improve monitoring
and inspection strategies used to secure the reliable operation of engineering structures.

%Expression \eqref{Damloc} gives a scalar
%characteristic of the damage localization process.
%This characteristic takes both damage and damage gradient into account to
%estimate the remaining lifetime of the structure.
%The influence of the damage localization on the strain localization
%is described in

%which is itself
%crucial for strain localization.

%Special thanks go to Prof. Altenbach and Dr. Naumenko for many
%useful suggestions.

\end{document}